\documentclass[reprint]{revtex4-2}

\usepackage{xcolor}

\usepackage[british]{babel}

\usepackage[ansinew]{inputenc} 
\usepackage{epsfig} 

\usepackage{graphicx}

\usepackage{xcolor}
\usepackage{physics}

\usepackage{amsmath,amssymb,amsthm,mathtools}
\usepackage{geometry}
\usepackage{microtype}
\usepackage[colorlinks=true,linkcolor=blue,citecolor=blue,urlcolor=blue]{hyperref}
\theoremstyle{remark}
\newtheorem*{remark}{Remark}

\usepackage{subfigure}
\usepackage{tikz}
\usepackage{pgf}
\usepackage{caption}
\usetikzlibrary{shapes,arrows,arrows.meta,positioning,calc}
\tikzset{
    line/.style = {
        draw,
        -{Latex[length=20pt, width=6pt]} 
    },
    cloud/.style = {
        draw, ellipse, node distance=2.5cm,
        minimum height=2em
    }
}

\begin{document}

\title{Time evolution of the Husimi and Glauber-Sudarshan functions in terms of complementary Hamiltonian symbols}
\author{Mritunjay Tyagi, Simon Friederich}
\email{m.tyagi@rug.nl, s.m.friederich@rug.nl}
\affiliation{University of Groningen, University College Groningen, Hoendiepskade 23/24, 9718BG Groningen, the Netherlands}

\begin{abstract}
We present a compact, systematic formulation of the dynamics of the Husimi Q- and Glauber-Sudarshan P-phase space distribution functions expressed in terms of their \emph{complementary} Hamiltonian symbols: Anti-Wick for Q and Wick for P. The resulting evolution equations have a universal leading structure, the classical Liouvillian drift plus terms with higher-order derivatives of the Hamiltonian. For Hamiltonians no higher than quartic in the moduli of the complex phase space variables $\alpha_i$, the higher-order terms reduce to a second-order Fokker-Planck type term with a \emph{traceless} diffusion matrix, thereby clarifying and recovering recent results for such Hamiltonians within a simple star-product framework. We further derive a transparent Ehrenfest theorem for Wick/Anti-Wick symbols of the operators representing dynamical observables. Using these results, we show that a previously reported nonclassical contribution to the Q-function drift for the anharmonic oscillator is an artifact of the quantization scheme used. Our paper consolidates the formulation of the dynamics of the phase space distribution functions using complementary symbols and provides an efficient route to compute and interpret quantum phase space evolution.        
\end{abstract}
\maketitle

\section{Introduction}

A quantum state can be expressed in phase space by a variety of quasi-probability functions, most famously the Wigner function $W$, the Husimi function $Q$, and the Glauber-Sudarshan function $P$ \cite{Wigner_1932,Husimi1940,ECG,Glauber,LEE1995147,folland}. All these functions are obtained from the density matrix $\hat\rho$ by subjecting it to some suitably chosen \emph{de-quantization} procedure, i.e. a reverse quantization mapping, from operators to phase space functions \cite{Ali:2004ft}, the operators' so-called \emph{symbols}. The Wigner function is obtained by reverse Weyl quantization, the Husimi function by reverse Wick quantization, and the Glauber-Sudarshan function by reverse Anti-Wick quantization. Correspondingly, these functions are referred to as, respectively, the Weyl, Wick (sometimes ``covariant'') and Anti-Wick (sometimes ``contravariant'') symbols \cite{folland, Ali:2004ft,Landsman1998} of the density matrix. Their differences notwithstanding, each of them encodes the same information content as the density matrix, and the density matrix can be recovered from any of them.

In this paper, we characterize the time evolution of the Husimi and Glauber-Sudarshan functions in terms of their \emph{complementary} Hamiltonian symbols, i.e. the time evolution of the Husimi function in terms of the Anti-Wick (contravariant) symbol of the Hamiltonian and the time evolution of the Glauber-Sudarshan function in terms of the Wick (covariant) symbol of the Hamiltonian. In the existing literature \cite{MIZRAHI1984241,MIZRAHI1986237,Appleby_2000,LEE1995147}, the time evolution of the Husimi and Glauber-Sudarshan functions is expressed in terms of Hamiltonian symbols of the same type -- Wick for the Q-function and Anti-Wick for the P-function. To our knowledge, a general formulation in terms of complementary Hamiltonian symbols has not been presented before. Although related expressions appear in special cases and for particular quantization choices, the complementary-symbol perspective as a systematic framework seems to be new. Our goal is to provide a compact and systematic formulation that clarifies the structure of phase-space dynamics and unifies existing results. We hope this repertoire will be of use in future applied and interpretative work.

In any case, studying the time evolution of the Husimi and Glauber-Sudarshan functions in terms of their \textit{complementary} Hamiltonian symbols, as done here, is worthwhile and natural. When one calculates quantum expectation values $\langle\hat A\rangle$ in phase space as weighted integrals $\int\, A(z)\,P(z) dz$, where $z$ is the phase space variable, $A(z)$ is some symbol associated with $\hat A$ and $P(z)$ an appropriately chosen quasi-probability distribution, one must always choose \emph{complementary} symbol types of $A$ and $P$ (where Weyl symbols are complementary to each other). Accordingly, the Hamiltonian symbol to naturally consider alongside the Husimi Q-function is the Anti-Wick Hamiltonian symbol, and the Hamiltonian symbol to naturally consider alongside the Glauber-Sudarshan P-function is the Wick Hamiltonian symbol. The fact that the time evolution of the Husimi  and Glauber-Sudarshan functions given in the literature so far is in terms of Hamiltonians of the same type seems to be mostly due to ease of derivation, not due to substantive physical reasons. \par

Here we close this gap and present evolution equations for the Husimi Q-  and Glauber-Sudarshan P-functions in terms of complementary Hamiltonian symbols. Our results turn out to have further benefits:
\begin{itemize}
\item In the time evolution of quasi-probability distributions expressed in terms of the complementary Hamiltonian the first two terms have the form of drift and diffusion terms of a Fokker-Planck equation, where the drift term corresponds to the classical Liouville evolution and the ``diffusion'' matrix has zero trace. The latter observation confirms a recent derivation of the time evolution of the Husimi Q-function for fourth-order Hamiltonians by \cite{drummond}, which we recover here elegantly using the formalism of star products and by clarifying the relation between the ``diffusion'' matrix and the second derivatives of the Hamiltonian. 
\item Based on the complementary quasi-probability distribution of an operator's symbol, an elegant expression of Ehrenfest's theorem can be derived for that symbol in terms of its associated (Wick or Anti-Wick) \emph{star product}. (See SECTION \ref{star products} for a review of star products and SECTION \ref{Ehrenfest theorem} for our generalized version of Ehrenfest's theorem.)
\item Our results allow for a transparent comparison between classical time evolution of the probability density in phase space and Husimi function in the quantum setting. Notably, they establish that an earlier finding by Milburn \cite{Milburn_article} according to which the ``drift'' part of the Husimi function time evolution in the anharmonic oscillator has a non-classical contribution arises from using a Hamiltonian not aligned with anti-Wick quantization. With the natural anti-Wick Hamiltonian, the classical Liouvillian drift is recovered and the dynamical consequences of the ``quantum drift'' term are revealed to be spurious.
\end{itemize}
The plan for this paper is as follows. In SECTION \ref{star products}, we review quantum phase space distribution functions and the results on their time evolution in terms of Hamiltonian symbols of the same type as found in the literature. 
In SECTION \ref{Q-function dynamics} we derive the time evolution of the Husimi Q-function using the Anti-Wick Hamiltonian symbol, in SECTION \ref{P-function dynamics} we derive the time evolution of the Glauber-Sudarshan P-function using the Wick Hamiltonian symbol. In SECTION \ref{Ehrenfest theorem}, we derive the Ehrenfest theorem for the evolution of expectation values of observables via the Wick and anti-Wick symbols of operators. In SECTION \ref{Anharmonic oscillator}, we consider Milburn's analysis of the anharmonic oscillator \cite{Milburn_article} and identify the appearance of a non-classical contribution to the drift in the evolution equation for the Husimi function to be an artifact of not considering the Anti-Wick quantized Hamiltonian, which is in fact the quantum Hamiltonian that is most naturally considered alongside use of the Husimi function. We conclude the paper in SECTION \ref{conclusion} with a brief summary.\\ \par

\section{$Q$- and $P$- function time evolution reviewed and re-derived}\label{star products}

\subsection{Phase space distribution functions and operator symbols}
In this section, we review and re-derive literature results \cite{MIZRAHI1984241, MIZRAHI1986237, Appleby_2000, LEE1995147} on the derivation of the time evolution equations of the $Q$- and $P$-functions. These derivations all describe the time evolution of the $Q$- and $P$-functions in terms of Hamiltonians of the same symbol-type, i.e. the Wick-symbol of the Hamiltonian for $Q$ and the Anti-Wick-symbol of the Hamiltonian for $P$. Our derivation is based on the star-products associated with the respective quantizations.

The different phase space distribution functions all contain the same information as the density matrix $\hat \rho(\hat a, \hat a^\dagger, t)$. The most well-known phase space distribution is the Wigner function, which can be calculated from the density matrix via:
\begin{equation}
    W(q,p,t)=2\int_{-\infty}^\infty e^{-2 i py/\hbar}\langle q+y|\hat \rho(\hat q, \hat p,t)|q-y\rangle\,dy\,. \label{wigner}
\end{equation}
The Husimi and Glauber-Sudarshan functions are introduced most conveniently in terms of a complex phase space variable $\alpha\in\mathbb{C} \cong\ \mathbb{R}^{2}$
\begin{equation}
    \alpha = \sqrt{\frac{m\omega}{2\hbar}}q+ i\sqrt{\frac{1}{2\hbar m \omega}}p\,
\end{equation}
with $m$ and $\omega$ chosen such that $\sqrt{\hbar/(m\omega)}$ is a characteristic length scale of the system, i.e. the oscillator length in case of the harmonic oscillator. For simplicity, we take $m\omega=1$.

The Husimi Q-function is defined in terms of coherent states $\ket{\alpha}$ centred at phase space locations $\alpha$
\begin{equation}
    Q(\alpha, \alpha^*,t)=\frac{1}{\pi}\bra{\alpha}\hat \rho(\hat a, \hat a^\dagger,t)\ket{\alpha}\,.\label{husimi}
\end{equation}
The Husimi Q-function is positive semi-definite, and it naturally emerges in heterodyne detection, where it represents the distribution of simultaneous measurements of conjugate quadratures \cite{PhysRevLett.117.070801}, but it also emerges in other setups \cite{Xu_2025}. It specifies the distribution of results in ``retrodictively optimal'' phase space measurement \cite{appleby1998optimaljointmeasurementsposition} and has recently been suggested as a key element in obtaining an ``ontological'' view of quantum theory \cite{Drummond_Reid,friederich2021}.

The Glauber-Sudarshan P-function, which is much less regular than the Husimi Q-function, can be specified implicitly, as the function $P(\alpha)$ that makes the density matrix diagonal in the basis $\lbrace\ket{\alpha}\rbrace$ of coherent states:
\begin{equation}
    \hat\rho=\int~d^2\alpha~ P(\alpha)\ket{\alpha}\bra{\alpha}\,.\label{glaubersudarshan}
\end{equation}

The Wigner, Husimi Q-, and Glauber-Sudarshan P-functions can be seen as the \emph{dequantized} versions of the density matrix in the Weyl, Wick, and Anti-Wick quantization schemes, respectively. For polynomial observables, Weyl, Wick and Anti-Wick quantization differ in that they assign phase space polynomials to symmetrically (Weyl), normally (Wick), and anti-normally ordered (Anti-Wick) operator polynomials (for the generalization of these quantization schemes to non-polynomial observables, see \cite{folland}).

The inverse of a quantization scheme is the corresponding ``dequantization.'' It involves mapping a self-adjoint operator $\hat A$ to a phase space function, its so-called ``symbol'', that is assigned to $\hat A$ under the chosen quantization scheme. The Weyl, Wick, and Anti-Wick dequantization, operator-to-symbol, maps are the following:

The Weyl symbol of a self-adjoint operator $\hat A$ is obtained by taking its Wigner transform (or Weyl de-quantization map $\mathcal{W}^{-1}$), which is given by
\begin{equation}
\begin{aligned}
    A_{Weyl}(q,p)=& \mathcal W^{-1}(\hat A(\hat q, \hat p))\,\\ 
    =&2\int_{-\infty}^\infty e^{-2 i py/\hbar}\langle x+y|\hat A|x-y\rangle\,dy\label{weyl}
\end{aligned}
\end{equation}
The Wick symbol of a self-adjoint operator $\hat A$, in turn, is obtained via the Wick de-quantization map $\tilde {\mathcal{W}}^{-1}$,
\begin{equation}
\begin{aligned}
A_{Wick}(\alpha,\alpha^*)=&\tilde{\mathcal W}^{-1}(\hat A(\hat a, \hat a^\dagger ))\,\\
=&\bra{\alpha}\hat A\ket{\alpha} \,,\label{wick}
\end{aligned}
\end{equation} 
where $\ket{\alpha}$ is the coherent state. Finally, the operator $\hat A$ has an anti-Wick symbol, obtained via anti-Wick dequatization map $\mathcal{AW}^{-1}$,
\begin{equation}
    A_{anti-Wick}(\alpha, \alpha^*)=\mathcal{AW}^{-1}(\hat A(\hat a, \hat a^\dagger))\,,
\end{equation}
if it can be written as
\begin{equation}
   \hat{A}(\hat a, \hat a^\dagger)=\int d^2\alpha ~A_{anti-Wick}(\alpha, \alpha^*)~\ket{\alpha}\bra{\alpha} \,. \label{anti-wick}
\end{equation}
\\ \par
A comparison between Eq.\ (\ref{wigner}) and Eq.\ (\ref{weyl}), between Eq.\ (\ref{husimi}) and Eq.\ (\ref{wick}), and between Eq.\ (\ref{glaubersudarshan}) and Eq.\ (\ref{anti-wick}) establishes the claim made earlier: that the Wigner, Husimi Q-, and Glauber-Sudarshan P-functions are, respectively, the Weyl, Wick, and Anti-Wick symbols of the density matrix.

The time evolution of the quantum state according to the Schr\"odinger equation for the state vector and the von Neumann equation for the density matrix translates into time evolutions for the Wigner, Husimi Q-, and Glauber-Sudarshan P-functions. These time evolutions are most conveniently calculated using the associated quantizations' \emph{star products}. For a given quantization scheme $\mathcal Q$ (not to be confused with the Husimi function $Q$), defined on a suitably chosen space of phase space functions, its non-commutative star product, which defines a deformed algebra on space of functions $A$, is defined as the operation $*$ that corresponds to the commutator at the level of Hilbert space operators:
\begin{equation}
\begin{aligned}
    [\mathcal Q(A),\mathcal Q(B)]=&-\frac{i}{\hbar}\mathcal{Q}(\{\{A,B\}\}_*)\,,\\
    =&-\frac{i}{\hbar} \mathcal{Q}(A*B-B*A)\,.\label{star-product}
\end{aligned}
\end{equation}
Star-products can be expressed as formal series with $\hbar$-dependent higher-order correction terms that make the star-products non-commutative \cite{Ali:2004ft,Landsman1998}. The most well-known star product is the Moyal star-product \cite{Moyal_1949}, which is associated with Weyl quantization. We provide the Wick and Anti-Wick star-products in Eqs.\ (\ref{wick-product}) and (\ref{anti-wick-product}), respectively. Groenewold's theorem \cite{GROENEWOLD1946405} implies that no quantization scheme satisfying natural conditions (linearity, correct mapping of basic variables, and commutator-Poisson correspondence) can reproduce the Poisson bracket for all observables. Equivalently, the star bracket $\{\{.,.\}\}_*$
cannot in general coincide with the Poisson bracket for arbitrary phase-space functions $A,B$ \cite{Ali:2004ft}. \\ \par

Importantly for the purposes of this article, if any of the Husimi Q- or Glauber-Sudarshan P-functions are used to calculate the quantum expectation value $\langle\hat A\rangle=Tr(\hat A\hat\rho)$ of an operator $\hat A$ as a weighted phase space integral, they must be used together with a \emph{complementary}, i.e. opposite-type, symbol of $\hat A$. More generally, if $\langle\hat A\rangle=Tr(\hat A\hat\rho)$ is calculated as a weighted phase space integral using some phase space distribution function $F_o(\alpha,\alpha^*,t)$, where $o$ denotes some dequantization scheme, then a symbol $A_{\bar o}$ of $\hat A$ must be used, where $\bar o$ denotes the complementary dequantization scheme.
\begin{equation}
\begin{aligned}
    \langle\hat A\rangle = & Tr(\hat A(\hat q, \hat p) \hat\rho(\hat q, \hat p, t))\,,\\
    = & \int ~d^2\alpha~A_{\bar o}F_o(t)\,, \label{complementary}
\end{aligned}
\end{equation}
Here Wick and Anti-Wick dequantization are complementary to each other, Weyl quantization is complementary to itself. (For a systematic review that also considers further ordering schemes, see \cite{LEE1995147}.) Consequently, if Wick/Anti-Wick dequantization is used, since the symbol-class of the observable operator and the density matrix are different, one cannot use any star-product to calculate the expectation values, unlike in the case of the Wigner function.
\\ \par 

\subsection{Dynamics of the distribution functions}
We briefly review the derivation of the time evolution of the Wigner, Husimi, and Glauber-Sudarshan functions using their associated (Weyl, Wick, and Anti-Wick) star-products. The time evolution of the Wigner function can be derived from the von-Neumann equation for the density matrix and expressed in terms of the Moyal star product $\{\{\cdot,\cdot\}\}_{*_{Weyl}}$ \cite{Moyal_1949}, see \cite{QM_phase_space_book} for further details:
\begin{equation}\label{Wigner dynamical equation}
\begin{aligned}
    \frac{\partial}{\partial t}W(q,p,t)&=\frac{\partial}{\partial t}\left(\mathcal{W}^{-1}\hat \rho(t)\right)\,,\\
    &=\mathcal{W}^{-1}\left(\frac{\partial}{\partial t}\hat \rho(t)\right)\,,\\
    &=-\frac{i}{\hbar}\left(\mathcal{W}^{-1}([\hat H, \hat{\rho}])\right)\,.\\
    &=-\frac{i}{\hbar}\{\{H_{Weyl},W(t)\}\}_{*_{Weyl}}\,.
\end{aligned}
\end{equation}
\\ \par
A derivation of the dynamical equations of the Husimi Q- and Glauber-Sudarshan P-functions is given by \cite{MIZRAHI1984241,MIZRAHI1986237}. These equations can be written as follows using the star-products of the respective dequantizations. As mentioned before, the Q-function is the Wick symbol of the density matrix. By Eq.\ (\ref{star-product}), the Wick symbol of an operator product $\hat A\cdot \hat B$ is given by $\bra{\alpha}\hat A\cdot \hat B\ket{\alpha}=A_{Wick}(\alpha,\alpha^*) \,*_{Wick}\, B_{Wick}(\alpha,\alpha^*)$ (and analogously for its Anti-Wick symbol). (In what follows, to keep notation simple, we drop the arguments $(\alpha,\alpha^*)$ of the phase space functions in the calculations ahead, and simply write $A_W$ and $A_{aW}$ for denoting Wick and Anti-Wick symbols of the operator $\hat{A}$, respectively (similarly for star products)). The time evolution equation for the Husimi function is obtained by taking the von-Neumann equation and sandwiching both sides between coherent state vectors:
\begin{equation}\label{dynamical equation for Q with wick symbols}
    \begin{aligned}
        \frac{\partial}{\partial t} Q(\alpha, \alpha^*,t) & =\bra{\alpha}\frac{\partial}{\partial t} \hat{\rho} (\hat a, \hat a^\dagger, t)\ket{\alpha}\,,\\
        & = -\frac{i}{\hbar}\bra{\alpha}[\hat H(\hat a, \hat{a}^\dagger), \hat{\rho} (\hat a, \hat a^\dagger, t)]\ket{\alpha}\,,\\
        & =-\frac{i}{\hbar} \bra{\alpha}\hat H \hat \rho \ket{\alpha} - \bra{\alpha}\hat \rho \hat H \ket{\alpha}\,,\\
        & =-\frac{i}{\hbar} H_{W} *_{W} Q - Q*_{W}H_{W}\,,\\
        & = -\frac{i}{\hbar}\{\{H_{W},Q\}\}*_{W}\,.
    \end{aligned}
\end{equation} 
\\ \par

Analogously, for the P-function, as the anti-Wick symbol of the density matrix, one has $\mathcal{AW}^{-1}(\hat A\cdot \hat B)=A_{aW}*_{aW}B_{aW}$. Proceeding as we did for the Wigner and Husimi Q-functions from the von-Neumann equation, we get,
\begin{equation}\label{dynamical equation for the P-function in terms of anti-Wick symbol}
\begin{aligned}
    \frac{\partial}{\partial t}P(t)&=\frac{\partial}{\partial t}\left(\mathcal{AW}^{-1}\hat \rho(t)\right)\,,\\
    &=\mathcal{AW}^{-1}\left(\frac{\partial}{\partial t}\hat \rho(t)\right)\,,\\
    &=-\frac{i}{\hbar}\left\{\mathcal{AW}^{-1}([\hat H, \hat{\rho}])\right\}\,,\\
    &=-\frac{i}{\hbar}\left\{(\mathcal{AW}^{-1}(\hat H\cdot \hat{\rho})-\mathcal{AW}^{-1}(\hat \rho \cdot \hat H)\right\}\,,\\
    &=-\frac{i}{\hbar}\left(H_{aW}*_{aW}P(t)-P(t)*_{aW} H_{aW}\right)\,,\\
    &=-\frac{i}{\hbar}\{\{H_{aW},P(t)\}\}_{*_{aW}}\,.
\end{aligned}
\end{equation}

\section{Q-function dynamics}\label{Q-function dynamics}

Eq.\ (\ref{dynamical equation for Q with wick symbols}) gives the dynamical equation for the Husimi function, itself the Wick symbol of the density matrix, in terms of the Wick symbol of the Hamiltonian. Here we derive its dynamical equation in terms of the \emph{Anti-Wick} symbol of the Hamiltonian, in many ways the more natural pairing, as evidenced by Eq.\ (\ref{complementary}). As we will show, using the anti-Wick symbol of the Hamiltonian the time evolution equation for the Husimi function takes the form
\begin{equation}\label{dynamical eqn for Q-function}
\resizebox{\columnwidth}{!}{%
    $\displaystyle
    \begin{aligned}
        \partial_tQ&=-\frac{i}{\hbar}\{\{H_{W}\,,Q\}\}_{*_{W}}\,,\\
    &=-\frac{i}{\hbar}\sum_{n=1}^{\infty}
    \frac{1}{n!}
    \;\left[\partial_{\alpha^*}^{n}\Bigl(
      \partial_{\alpha}^{n}H_{aW}\cdot Q
    \Bigr)-\partial_{\alpha}^{n}\Bigl(
       \partial_{\alpha^*}^{n}H_{aW}\cdot Q
    \Bigr)\right]\,.
    \end{aligned}
    $}
\end{equation}

We obtain this result from the following expression for the Wick star bracket in terms of an Anti-Wick symbol of one operator and the Wick symbol of the other operator:
\begin{equation}\label{Wick star bracket in terms of anti-Wick symbols}
\resizebox{\columnwidth}{!}{%
    $\displaystyle
\begin{aligned}
    \{\{f,g\}\}_{*_{W}}&=\sum_{n=1}^{\infty}
    \frac{1}{n!}
    \;\left[\partial_{\alpha^*}^{n}\!\Bigl(
      \partial_{\alpha}^{n}\,\bar f\;\cdot g
    \Bigr)-\partial_{\alpha}^{n}\!\Bigl(
      \bar \partial_{\alpha^*}^{n}\,\bar f\;\cdot g
    \Bigr)\right]\,.
\end{aligned}
$}
\end{equation}
Here $\bar f$ is the Anti-Wick symbol of the operator $\hat{f}$ which is related to the Wick symbol $f$ via $\bar f=e^{-\partial_{\alpha^*}\partial_{\alpha}}f$ \cite{Ali:2004ft}.\\ \par

We now derive this equation. Consider the phase space functions $f(\alpha,\alpha^*)$ and $g(\alpha,\alpha^*)$ (we drop the arguments in the calculations ahead) as the Wick symbols of operators $\hat f$ and $\hat{g}$. The Wick star product is given as \cite{MIZRAHI1984241},
\begin{equation}
\begin{aligned}
    f*_{W}g&=f e^{\overset{\xleftarrow{}}{\partial_{\alpha}}\overset{\xrightarrow{}}{\partial_{\alpha^*}}}g\,,\\
    &=\sum_{k=0}^\infty \frac{1}{k!} \partial_\alpha^k f \partial_{\alpha^*}^k g \,. \label{wick-product}
\end{aligned}
\end{equation}
where $\partial_\alpha=\sqrt{\frac{\hbar}{2}}(\partial_q-i\partial_p)$ (for any variable $s$ we denote $\partial_s\equiv \frac{\partial}{\partial_s}$) and $\partial_{\alpha^*}=(\partial_\alpha)^*$.

Next we use the result, demonstrated in Appendix \ref{Proof of identity used to shift derivatives}, that 
\begin{equation} \label{Shifting the derivatives to surface derivatives}
    \partial_{\alpha}^k f  \partial_{\alpha^*}^k g = \sum_{m=0}^k \left(-1\right)^{k-m} \, \binom{k}{m}\,  \partial_{\alpha^*}^{m}\left( \partial_{\alpha^*}^{k-m}\partial_{\alpha}^k f\, g\right)\,,
\end{equation}
where $\binom{n}{k}=\frac{n!}{k!(n-k)!}$. The Wick star product between $f$ and $g$ can be written as,
\begin{equation}
\resizebox{\columnwidth}{!}{%
    $\displaystyle
    f*_{W}g = \sum_{k=0}^\infty \frac{1}{k!}\left[\sum_{m=0}^k (-1)^{k-m}\, \binom{k}{m} \,  \partial_{\alpha^*}^m\left( \partial_{\alpha^*}^{k-m} \partial_{\alpha}^kf\, g\right)\right]\,.
    $}
\end{equation}
From the above we can also write down $g*_Wf$ by replacing $g \xleftrightarrow[]{}f$. It is clear from the above expressions that for the Wick star bracket $\{\{f,g\}\}_{*_{W}}=f*_{W}g-g*_{W}f$, $k=0$ terms will cancel with each other and for each k, $m=0$ terms from both star products will also cancel each other. 
So we get,
\begin{equation}\label{Wick star bracket as a sum in terms of Wick symbols}
\resizebox{\columnwidth}{!}{%
    $\displaystyle
\begin{aligned}
    \{\{f,g\}\}_{*_{W}} & = f*_{W}g-g*_{W}f\,,\\
    & = \sum_{k=1}^\infty \frac{1}{k!} \left[\sum_{m=1}^k (-1)^{k-m}\,\binom{k}{m}\right. \,\\
    & \,\left. \left\{\partial_{\alpha^*}^m\left(\partial_{\alpha^*}^{k-m} \partial_{\alpha}^k f\, g\right)-\partial_{\alpha}^m\left( \partial_{\alpha}^{k-m} \partial_{\alpha^*}^kf\, g\right)\right\}\right]\,.
\end{aligned}
$}
\end{equation}
Eq.\ (\ref{Wick star bracket in terms of anti-Wick symbols}) is obtained from this, as shown in Appendix \ref{Appendix B}.
 \\ \par

Having shown that one can express the Wick star bracket between two Wick-symbols $f$ and $g$ as a series with derivatives acting on products of derivatives of the anti-Wick symbol of one and the Wick symbol of the other we obtain for the dynamical equation for the Q-function, as given in equation (\ref{dynamical eqn for Q-function}):
\begin{equation*}
\resizebox{\columnwidth}{!}{%
    $\displaystyle
    \begin{aligned}
        \partial_tQ
    &=-\frac{i}{\hbar}\sum_{n=1}^{\infty}
    \frac{1}{n!}
    \;\left[\partial_{\alpha^*}^{n}\Bigl(
      \partial_{\alpha}^{n}H_{aW}\cdot Q
    \Bigr)-\partial_{\alpha}^{n}\Bigl(
       \partial_{\alpha^*}^{n}H_{aW}\cdot Q
    \Bigr)\right]\,.
    \end{aligned}
    $}
\end{equation*}
In a context where the Hamilton operator $\hat H$ has been obtained from a classical Hamiltonian $H$ via anti-Wick quantization (or coherent state quantization), $H_{aW}$ matches the classical Hamiltonian and one can simply write $H$ instead of $H_{aW}$. For Hamiltonians that can be written as functions of $|\alpha|$ and have no higher than quartic terms in $|\alpha|$, the above equation reduces to a Fokker-Planck equation, where $n=1$ term gives the drift contribution while $n=2$ term gives a diffusion-like contribution, which we discuss further below. For Hamiltonians of higher than quartic order in $|\alpha|$, we get additional contributions from the $n=3,4,...$ terms and the dynamical equation for the Q-function ceases to be a Fokker-Planck equation.  \\ \par

We can also write down the dynamical equation for the Q-function on the real phase space instead of the complex phase space. From the expressions for $\partial_{\alpha}$ and $\partial_{\alpha^*}$ and using the binomial expansion, we get,
\begin{equation}\label{n-order complex phase space derivative in terms of real phase space derivatives}
\begin{aligned}
    \partial_{\alpha}^n
=& \left(\frac{\hbar}{2}\right)^{n/2}\sum_{m=0}^n\binom{n}{m}(-i)^m\,\partial_q^{\,n-m}\partial_p^m\,,\\
\partial_{\alpha^*}^n
=&\left(\frac{\hbar}{2}\right)^{n/2}\sum_{k=0}^n\binom{n}{k} i^k\,\partial_q^{\,n-k}\partial_p^k.
\end{aligned}
\end{equation}
Plugging into,
\begin{equation*}
\begin{aligned}
   \{\{f,g\}\}_{*_{W}} =& \sum_{n=1}^\infty\frac1{n!}\Bigl[
\partial_{\alpha^*}^n\bigl(\partial_{\alpha}^n\bar f\;\cdot g\bigr)
-
\partial_{\alpha}^n\bigl(\partial_{\alpha^*}^n\bar f\;\cdot g\bigr)
\Bigr]  \,,\\
=& \sum_{n=1}^\infty\frac1{n!} \left(\frac{\hbar}{2}\right)^n
\sum_{k=0}^n\sum_{m=0}^n
\binom{n}{k}\binom{n}{m}\\
& ~~\bigl[i^k(-i)^m-(-i)^k i^m\bigr]\;\\&~~~~~~~~~~~~~~~~
\cdot \partial_q^{\,n-k}\partial_p^{\,k}
\Bigl[
\partial_q^{\,n-m}\partial_p^{\,m}\bar f\;\cdot g
\Bigr].
\end{aligned}    
\end{equation*}
which is a double sum over $k,m=0,\dots,n$. In this form, each term is exactly of the kind,
\begin{equation*}
\partial_q^{\,n-k}\partial_p^{\,k}
\bigl(\partial_q^{\,n-m}\partial_p^m\,\bar f\;\cdot g\bigr),
\end{equation*}
and the overall coefficient is
$C^n_{k,m}=\frac1{n!}\binom nk\binom nm\bigl[i^k(-i)^m-(-i)^k i^m\bigr]$. We get pure $q$-surface derivatives for $k=0$, and pure $p$-surface derivatives for $k=n$.\\ \par

From these expressions we can write down the dynamical equation for the Q-function in terms of the real phase space variables,
\begin{equation*}
    \begin{aligned}
        \partial_tQ=&-\frac{i}{\hbar}\{\{H_{W}\,,Q\}\}_{*_{W}}\,,\\
         =& -\frac{i}{\hbar}\sum_{n=1}^\infty \left(\frac{\hbar}{2}\right)^n\\
&~~~~\sum_{k,m=0}^n
C^n_{k,m}\;
\partial_q^{\,n-k}\partial_p^{\,k}
\Bigl[
\partial_q^{\,n-m}\partial_p^{\,m}H_{aW}\;\cdot Q
\Bigr]\,.
    \end{aligned}
\end{equation*}
Restricting to Hamiltonians of at most quartic order in $|\alpha|$ so that $\partial_\alpha^n H= \partial_{\alpha^*}^n H=0$ for $n\geq3$, we get the Fokker-Planck equation for the Q-function (in terms of the real phase space variables),
\begin{equation*}
    \begin{aligned}
        \partial_tQ=& -\Bigr[\partial_q(A_q \cdot Q)+\partial_p(A_p \cdot Q)\Bigl]\\
        & +\frac{1}{2}\Bigr[\partial_q^2(D_{qq}\cdot Q)+\partial_p^2(D_{pp}\cdot Q)\\
        &~~~~~~~~+\partial_{qp}((D_{pq}+D_{qp})\cdot Q)\Bigl]\,.
    \end{aligned}
\end{equation*}
The drift ($n=1$) and diffusion $(n=2)$ terms are given as,
\begin{equation}
    \begin{aligned}
        A_q&=\partial_pH_{aW}\,,\\
        A_p&=-\partial_qH_{aW}\,,\\
        D_{pp}&=\hbar \partial_q\partial_pH_{aW}\,,\\
        D_{qq}&=-\hbar \partial_q\partial_pH_{aW}=-D_{pp} \,,\\
        D_{qp}&=D_{pq}=\frac{1}{2}\hbar(\partial_q^2 -\partial_p^2)H_{aW}\,.
    \end{aligned}\label{drift-diffusion}
\end{equation}

The Fokker-Planck equation for the Q-function becomes
\begin{equation}
    \begin{aligned}\label{Fokker-Planck}
        \partial_tQ 
        =& -\left[\partial_q\left(\partial_pH_{aW} \cdot Q\right)-\partial_p\left(\partial_q H_{aW} \cdot Q\right)\right]\\
        & +\frac{\hbar}{2}\left[(\partial_p^2-\partial_q^2)\left(\partial_q\partial_p  H_{aW}\cdot Q\right)\right.\\
        &\left.~~~~~~~~+\partial_q\partial_p\left((\partial_q^2-\partial_p^2) H_{aW}\cdot Q\right)\right]\,.
    \end{aligned}
\end{equation}
This result reproduces an observation by \cite{drummond}, that the time evolution of the Husimi function in theories with Hamiltonians up to fourth order in $|\alpha|$ has the form of a zero-trace, $D_{qq}+D_{pp}=0$, diffusion. The vanishing trace of the diffusion matrix indicates that the quantum corrections modify the flow without introducing net contraction or expansion in phase space volume. This observation highlights the structural analogy with Liouvillian flow, while keeping the quantum corrections distinct from classical stochastic diffusion. A fuller discussion of interpretational implications lies beyond our scope.

For Hamiltonians of higher order in $|\alpha|$, there are higher order beyond diffusion corrections. The first beyond diffusion terms are obtained for $n=3$ which are the third order surface derivatives, 
    \begin{equation*}
    \resizebox{\columnwidth}{!}{%
    $\displaystyle
    \begin{aligned}
        -\frac{i}{\hbar}\{\{H_{W},Q\}\}_{*_{W}}^{(n=3)}
        = \frac{\hbar^2}{3! 2^2}\Bigl[&\left(\partial_q^3-3\partial_q\partial_p^2\right)\left(\left(\partial_p^3-3\partial_q^2\partial_p\right)H\cdot Q\right)\\
        &-\left(\partial_p^3-3\partial_q^2\partial_p\right)\left(\left(\partial_q^3-3\partial_q\partial_p^2\right)H \cdot Q\right)\Bigr]\,.
    \end{aligned}
    $}
\end{equation*}\\ \par

The calculation above for the single mode can be extended to $N-$modes (real phase space dimension $2N$).
\begin{equation}
\begin{aligned}
\mathbf{m}=&(m_1,\dots,m_N)\in\mathbb{N}^N\,,\\\qquad
|\mathbf{m}|=&\sum_{j=1}^N m_j\,,\\\qquad
\mathbf{m}!:=&\prod_{j=1}^N m_j!\,.
\end{aligned}
\end{equation}
and for derivatives, write
\begin{equation}
\partial^{\mathbf{m}}:=\prod_{j=1}^N \partial_{\alpha_j}^{\,m_j},\qquad
\bar\partial^{\mathbf{m}}:=\prod_{j=1}^N \partial_{\alpha_j^*}^{\,m_j}\,.
\end{equation}

The multi-mode Wick star product is
\begin{equation}
\begin{aligned}
f*_{W}g=& f\,\exp\!\Bigl(\sum_{j=1}^N\overset{\xleftarrow{}}{\partial_{\alpha_j}}\,
\overset{\xrightarrow{}}{\partial_{\alpha_j^*}}\Bigr)\,g\\
=& \sum_{\mathbf{k}\in\mathbb{N}^N}\frac{1}{\mathbf{k}!}\,
\partial^{\mathbf{k}}f\;\bar\partial^{\mathbf{k}}g \,.
\end{aligned}
\end{equation}

Equation (\ref{Wick star bracket in terms of anti-Wick symbols}) for $N-$modes is
\begin{equation}
\{\{f,g\}\}_{*_{W}}
=\sum_{|\mathbf{m}|\ge1}\frac{1}{\mathbf{m}!}
\Big[
\bar\partial^{\mathbf{m}}\!\bigl(\partial^{\mathbf{m}}\bar f\cdot g\bigr)
-
\partial^{\mathbf{m}}\!\bigl(\bar\partial^{\mathbf{m}}\bar f\cdot g\bigr)
\Big],
\end{equation}
where $\bar f=\exp\Bigl(-\sum_{j=1}^N\partial_{\alpha_j}\partial_{\alpha^*_j}\Bigr)f$. The dynamical equation for the Q-function dynamics reads
\begin{equation} \resizebox{\columnwidth}{!}{%
    $\displaystyle
\begin{aligned}
\partial_t Q &= -\frac{i}{\hbar}\{\{H_{W},Q\}\}_{*_{W}}\\
& = -\frac{i}{\hbar}\sum_{|\mathbf{m}|\ge1}\frac{1}{\mathbf{m}!}
\Big[
\bar\partial^{\mathbf{m}}\!\bigl(\partial^{\mathbf{m}} H_{aW}\cdot Q\bigr)
-
\partial^{\mathbf{m}}\!\bigl(\bar\partial^{\mathbf{m}} H_{aW}\cdot Q\bigr)
\Big]\,.
\end{aligned}$}
\end{equation}
To pass to the real phase space, for each mode $j$, 
\begin{equation}
\begin{aligned}
\partial_{\alpha_j}=\sqrt{\tfrac{\hbar}{2}}\;(\partial_{q_j}-i\partial_{p_j})\,,\\
\partial_{\alpha_j^*}=\sqrt{\tfrac{\hbar}{2}}\;(\partial_{q_j}+i\partial_{p_j}).
\end{aligned}
\end{equation}

For a multi-index \(\mathbf{m}\) the operator \(\partial^{\mathbf{m}}\) expands as the product of one-mode binomial expansions
\begin{equation}
\partial^{\mathbf{m}}
= \bigl(\tfrac{\hbar}{2}\bigr)^{|\mathbf{m}|/2}\prod_{j=1}^N\left(\sum_{u_j=0}^{m_j}\binom{m_j}{u_j}(-i)^{u_j}
\partial_{q_j}^{\,m_j-u_j}\partial_{p_j}^{\,u_j}\right),
\end{equation}
and similarly for \(\bar\partial^{\mathbf{m}}\). In terms of real phase space coordinates, the dynamical equation for the Q-function is
\begin{equation}
\begin{aligned}
\partial_t Q(t)
&= -\frac{i}{\hbar}\sum_{|\mathbf{m}|\ge1}\frac{1}{\mathbf{m}!}\Bigl(\tfrac{\hbar}{2}\Bigr)^{|\mathbf{m}|}\\ &
\sum_{0\le \mathbf u, \mathbf v\le \mathbf m}
\Biggl(\prod_{j=1}^N \binom{m_j}{u_j}\binom{m_j}{v_j}\Biggr)
\Bigl(i^{\,|\mathbf{v}|}(-i)^{\,|\mathbf{u}|}\Bigr)\\
&
\Biggl\{
\partial_{q}^{\,\mathbf{m}-\mathbf{v}}\partial_{p}^{\,\mathbf{v}}\Big[
\partial_{q}^{\,\mathbf{m}-\mathbf{u}}\partial_{p}^{\,\mathbf{u}} H_{aW}\cdot Q(t)
\Big]\\
&-
\partial_{q}^{\,\mathbf{m}-\mathbf{u}}\partial_{p}^{\,\mathbf{u}}\Big[
\partial_{q}^{\,\mathbf{m}-\mathbf{v}}\partial_{p}^{\,\mathbf{v}} H_{aW}\cdot Q(t)
\Big]
\Biggr\}.
\end{aligned}
\end{equation}

\section{P-function dynamics}\label{P-function dynamics}
In this section we express the dynamical equation for the P-function (with Anti-Wick symbol of the Hamiltonian) i.e., equation (\ref{dynamical equation for the P-function in terms of anti-Wick symbol}) in terms of the Wick symbol of the Hamiltonian operator. The expression is given as,
\begin{equation}\label{Dynamical equation for P-function}\resizebox{\columnwidth}{!}{%
    $\displaystyle
    \begin{aligned}
        \partial_tP&=-\frac{i}{\hbar}\{\{H_{aW}\,,P\}\}_{*_{aW}}\,,\\
    &=\frac{i}{\hbar}\sum_{n=1}^{\infty}
    \frac{(-1)^n}{n!}\left[\partial_{\alpha^*}^{n}\Bigl(
      \partial_{\alpha}^{n}\,H_{W}\cdot P
    \Bigr)-\partial_{\alpha}^{n}\Bigl(
       \partial_{\alpha^*}^{n}H_{W}\cdot P
    \Bigr)\right]\,.
    \end{aligned}$}
\end{equation}
\\ \par 

To this end, like in the case of the Q-function, we derive a general expression for the Anti-Wick star bracket in terms of the Wick symbol of one operator and Anti-Wick symbol of the other operator. We now consider $f(\alpha,\alpha^*)$ and $g(\alpha,\alpha^*)$ as the Anti-Wick symbols of the operators $\hat f$ and $\hat g$ (unlike the previous section where they were used to denote the Wick symbols). The Anti-Wick star product is given as \cite{MIZRAHI1984241,MIZRAHI1986237},
\begin{equation}
\begin{aligned}
    f*_{aW}g&=f e^{-\overset{\xleftarrow{}}{\partial_{\alpha^*}}\overset{\xrightarrow{}}{\partial_{\alpha}}}g\,,\\
    &=\sum_{k=0}^\infty \frac{(-1)^k}{k!} \partial_{\alpha^*}^k f  \partial_{\alpha}^k g \,. \label{anti-wick-product}
\end{aligned}
\end{equation}
Using equation (\ref{Shifting the derivatives to surface derivatives}), we can write,
\begin{equation}
\begin{aligned}
    \partial_{\alpha^*}^k f  \partial_{\alpha}^k g = \sum_{m=0}^k \left(-1\right)^{k-m} \binom{k}{m}\,\\  \partial_{\alpha}^{m}\left( \partial_{\alpha}^{\,k-m}\partial_{\alpha^*}^k f\, g\right)\,.
\end{aligned}
\end{equation}
The Anti-Wick star product between $f$ and $g$ can now be written as,
\begin{equation}
\begin{aligned}
    f*_{aW}g = \sum_{k=0}^\infty \frac{(-1)^k}{k!}\left[\sum_{m=0}^k (-1)^{k-m}\binom{k}{m} \right.\\ \left.\partial_{\alpha}^m\left( \partial_{\alpha}^{\,k-m} \partial_{\alpha^*}^kf\, g\right)\right]\,.
\end{aligned}
\end{equation}
From the above we can also write down $g *_{aW}f$ by replacing $g \xleftrightarrow[]{}f$. It is clear from the above expression that just like the Wick star bracket, for the Anti-Wick star bracket $f*_{aW}g-g*_{aW}f$, the $k=0$ terms will cancel with each other and for each k, $m=0$ terms from both star products will also cancel each other.
So we get
\begin{equation*}\resizebox{\columnwidth}{!}{%
    $\displaystyle
\begin{aligned}
    \{\{f,g\}\}_{*_{aW}} & = f*_{aW}g-g*_{aW}f\,,\\
    & = - \sum_{k=1}^\infty \frac{(-1)^k}{k!} \left[\sum_{m=1}^k(-1)^{k-m} \binom{k}{m}\right.\,\\
    &~~~~\left.\left\{\partial_{\alpha^*}^m\left(\partial_{\alpha^*}^{\,k-m} \partial_{\alpha}^kf\, g\right)-\partial_{\alpha}^{m}\left( \partial_{\alpha}^{\,k-m} \partial_{\alpha^*}^kf\, g\right)\right\}\right]\,.
\end{aligned}$}
\end{equation*}
This expression above looks structurally similar to the equation (\ref{Wick star bracket as a sum in terms of Wick symbols}), with the difference being an overall negative sign and having the term $\frac{(-1)^k}{k!}$ instead of $\frac{1}{k!}$ in the $k$-sum. Following the same steps as presented in Appendix \ref{Appendix B}, we get,
\begin{equation}\label{anti-Wick star product in terms of the Wick symbols}
\begin{aligned}
    &\{\{f,g\}\}_{*_{aW}}\\&=-\sum_{n=1}^{\infty}
    \frac{(-1)^n}{n!}
    \;\left[\partial_{\alpha^*}^{n}\!\Bigl(
      \partial_{\alpha}^{n}\,\tilde f\;\cdot g
    \Bigr)-\partial_{\alpha}^{n}\!\Bigl(
      \partial_{\alpha^*}^{n}\,\tilde f\;\cdot g
    \Bigr)\right]\,,
\end{aligned}
\end{equation}
where $\tilde f=e^{\frac{\hbar}{2}\partial_{\alpha^*}\partial_{\alpha}}f$ is the Wick symbol obtained via the Berezin transform of the Anti-Wick symbol $f$ \cite{Ali:2004ft}.

Using the above expression for the Anti-Wick star bracket, we obtain equation (\ref{Dynamical equation for P-function}) the dynamical equation for the P-function in terms of the Wick symbol of the Hamiltonian,
\begin{equation*}\resizebox{\columnwidth}{!}{%
    $\displaystyle
    \begin{aligned}
        \partial_tP&=-\frac{i}{\hbar}\{\{H_{aW}\,,P\}\}_{*_{aW}}\,,\\
    &=\frac{i}{\hbar}\sum_{n=1}^{\infty}
    \frac{(-1)^n}{n!}\\
    &~~~~~~~~~~~~~~~~~~\left[\partial_{\alpha^*}^{n}\Bigl(
      \partial_{\alpha}^{n}\,H_{W}\cdot P
    \Bigr)-\partial_{\alpha}^{n}\Bigl(
       \partial_{\alpha^*}^{n}H_{W}\cdot P
    \Bigr)\right]\,.
    \end{aligned}$}
\end{equation*}

Here again for a special class of Hamiltonians i.e. Hamiltonians that are upto quartic order in $|\alpha|$, the equation above reduces to the Fokker-Planck equation where $n=1$ term gives the drift contribution while $n=2$ term gives the diffusion contribution. 
For Hamiltonians of higher than quartic order in $|\alpha|$, we get additional contributions from the $n=3,4,...$ terms and the dynamical equation for the P-function, just like for Q-function, ceases to be a Fokker-Planck equation.   \par

The dynamical equation for the P-function in terms of the real phase space variables can be calculated following the same steps as for the Q-function. Comparing the dynamical equation for the P- and Q-functions (equation (\ref{dynamical eqn for Q-function}) and equation (\ref{Dynamical equation for P-function})), it can be easily seen that they differ by an overall negative sign and also one expression has $\frac{1}{n!}$ in the $n$-sum while the other has $\frac{(-1)^n}{n!}$. So accounting for this difference and following the same steps as we did for the case of Q-functions, i.e., plugging in equation (\ref{n-order complex phase space derivative in terms of real phase space derivatives}) in the expression for Anti-wick star bracket and simplifying, we get,
\begin{equation*}
    \begin{aligned}
        \partial_tP&=-\frac{i}{\hbar}\{\{H_{aW}\,,P\}\}_{*_{aW}}\,,\\
        & = \frac{i}{\hbar}\sum_{n=1}^\infty \left(\frac{\hbar}{2}\right)^n
\sum_{k,m=0}^n
C^n_{k,m}\\
&~~~~~~~~~\partial_q^{\,n-k}\partial_p^{\,k}
\Bigl[
\partial_q^{\,n-m}\partial_p^{\,m}H_{W}\;\cdot P
\Bigr]\,,
    \end{aligned}
\end{equation*}
where $C^n_{k,m}=\frac{(-1)^n}{n!}\binom nk\binom nm\bigl[i^k(-i)^m-(-i)^k i^m\bigr]$. \\ \par  

The Fokker-Planck equation (in terms of real phase space variables) for Hamiltonians where $\partial_{\alpha}^n H_{W}=\partial_{\alpha^*}^n H_{W}=0$ for $n\geq3$ is,
\begin{equation*}\resizebox{\columnwidth}{!}{%
    $\displaystyle
    \begin{aligned}
        \partial_tP=& -\Bigr[\partial_q(A_q \cdot P)+\partial_p(A_p \cdot P)\Bigl]\\
        &+\frac{1}{2}\Bigr[\partial_q^2(D_{qq}\cdot P)+\partial_p^2(D_{pp}\cdot P)+\partial_{qp}((D_{pq}+D_{qp})\cdot P)\Bigl]\,.
    \end{aligned}$}
\end{equation*}
with the drift ($A_q, A_p$) and diffusion $(D_{qq}, D_{pp}$ and $D_{pq})$ terms given as,
\begin{equation*}
    \begin{aligned}
        A_q&=\partial_pH_{W}\,,\\
        A_p&=-\partial_qH_{W}\,,\\
        D_{pp}&=-\hbar \partial_q\partial_pH_{W}\,,\\
        D_{qq}&=\hbar \partial_q\partial_pH_{W}=-D_{pp} \,,\\
        D_{qp}&=D_{pq}=\frac{1}{2}\hbar(\partial_p^2 -\partial_q^2)H_{W}\,.
    \end{aligned}
\end{equation*}

\emph{Remark:} In the case of the Wigner function, where the dynamical equation is given by equation (\ref{Wigner dynamical equation}), from the structure of the Moyal star bracket (which is expressed as a sine series of phase space derivatives) one can infer that all the even ordered derivatives in the expression are absent. This implies that for the cases where one can reduce the dynamical equation for the Wigner function to a Fokker-Planck equation, the diffusion always vanishes and in that sense the dynamics are always deterministic.  \\
\par 

For $N-$modes, the equation ($\ref{anti-Wick star product in terms of the Wick symbols}$) can be generalized to
\begin{equation}\resizebox{\columnwidth}{!}{%
    $\displaystyle
\{\{f,g\}\}_{*_{aW}}
= -\sum_{|\mathbf{m}|\ge1}\frac{(-1)^{|\mathbf{m}|}}{\mathbf{m}!}
\Big[
\bar\partial^{\mathbf{m}}\!\bigl(\partial^{\mathbf{m}}\widetilde f\cdot g\bigr)
- \partial^{\mathbf{m}}\!\bigl(\bar\partial^{\mathbf{m}}\widetilde f\cdot g\bigr)
\Big]\,,$}
\end{equation}
where $\tilde f=\exp\Bigl(\sum_{j=1}^N\partial_{\alpha_j}\partial_{\alpha^*_j}\Bigr)f$. 
Using this, the dynamical equation for the P-function is 
\begin{equation}\resizebox{\columnwidth}{!}{%
    $\displaystyle
\begin{aligned}
\partial_t P&= -\frac{i}{\hbar}\{\{H_{aW},P\}\}_{*_{aW}}\\
&= \frac{i}{\hbar}\sum_{|\mathbf{m}|\ge1}\frac{(-1)^{|\mathbf{m}|}}{\mathbf{m}!}
\Big[
\bar\partial^{\mathbf{m}}\!\bigl(\partial^{\mathbf{m}}H_{W}\cdot P\bigr)
- \partial^{\mathbf{m}}\!\bigl(\bar\partial^{\mathbf{m}}H_{W}\cdot P\bigr)
\Big]\,.
\end{aligned}$}
\end{equation}
In real phase space coordinates, the above dynamical equation is written as,
\begin{equation}\resizebox{\columnwidth}{!}{%
    $\displaystyle
\begin{aligned}
\partial_t P(q,p,t)
&= \frac{i}{\hbar}\sum_{|\mathbf{m}|\ge1}\frac{(-1)^{|\mathbf{m}|}}{\mathbf{m}!}\Bigl(\tfrac{\hbar}{2}\Bigr)^{|\mathbf{m}|}\\
&\sum_{0\le \mathbf u, \mathbf v\le \mathbf m}
\Biggl(\prod_{j=1}^n \binom{m_j}{u_j}\binom{m_j}{v_j}\Biggr)
\Bigl(i^{\,|\mathbf{v}|}(-i)^{\,|\mathbf{u}|}\Bigr)\\
&
\Biggl\{
\partial_{q}^{\,\mathbf{m}-\mathbf{v}}\partial_{p}^{\,\mathbf{v}}\Big[
\partial_{q}^{\,\mathbf{m}-\mathbf{u}}\partial_{p}^{\,\mathbf{u}}H_{W}(q,p)\;\cdot\; P(q,p,t)
\Big]\\
&-\partial_{q}^{\,\mathbf{m}-\mathbf{u}}\partial_{p}^{\,\mathbf{u}}\Big[
\partial_{q}^{\,\mathbf{m}-\mathbf{v}}\partial_{p}^{\,\mathbf{v}}H_{W}(q,p)\;\cdot\; P(q,p,t)
\Big]
\Biggr\}.
\end{aligned}$}
\end{equation}

\section{Ehrenfest Theorem}\label{Ehrenfest theorem}

A version of the Ehrenfest theorem for the evolution of expectation values can be written in terms of Weyl symbols and the Moyal bracket:
\begin{equation}
    \begin{aligned}
        \frac{d}{dt}\langle\hat A\rangle 
        & = -\frac{i}{\hbar}\langle\{\{A_{Weyl}\,,\,H_{Weyl}\}\}_{*_{Weyl}}\rangle\,.\label{ehrenfest_wigner}
    \end{aligned}
\end{equation}
This equation can be derived straightforwardly using the Moyal bracket 
and the cyclic property of the Moyal star product under the integral
(Eqs. (18) and (50) in\cite{QM_phase_space_book}): 
\begin{equation*}\resizebox{\columnwidth}{!}{%
    $\displaystyle
    \begin{aligned}
        \frac{d}{dt}\langle\hat A\rangle & =\frac{d}{dt}\Bigl(\int~dq~dp~A_{Weyl}W(t) \Bigr)\,,\\
        & = \frac{d}{dt}\Bigl(\int~dq~dp~A_{Weyl}*_{Weyl}W(t) \Bigr)\,,\\
        &=\int~dq~dp~A_{Weyl}*_{Weyl}\Bigl(\frac{d}{dt}W(t)\Bigr)\,,\\
        &=-\frac{i}{\hbar}\int~dq~dp~A_{Weyl}*_{Weyl}\{\{H_{Weyl},W(t)\}\}_{*_{Weyl}}\,,\\
        &=-\frac{i}{\hbar}\int~dq~dp~\{\{A_{Weyl},H_{Weyl}\}\}_{*_{Weyl}} *_{Weyl} W(t)\,,\\
        & = -\frac{i}{\hbar}\langle\{\{A_{Weyl}\,,\,H_{Weyl}\}\}_{*_{Weyl}}\rangle\,.
    \end{aligned}$}
\end{equation*}
Versions of the Ehrenfest theorem analogous to Eq.\ (\ref{ehrenfest_wigner}), but with Wick and Anti-Wick symbols in place of Weyl symbols, can be given. However, their derivation is more complicated than the above derivation for Weyl symbols. The reason for this is that in the derivation of the Ehrenfest theorem in terms of Weyl symbols, all symbols that appear are Weyl symbol-class and, so, only the Moyal star-product must be used. In the derivation for Wick (anti-Wick) symbols, the phase space distribution to be used in analogy to the Wigner function is the Glauber-Sudarshan P-(Husimi Q-) function, which is not of the same symbol-type, but rather of the complementary one, leading to an algebraic mismatch. 

However, the results derived in Eq.\ (\ref{dynamical eqn for Q-function}) for the Q-function and Eq.\ (\ref{Dynamical equation for P-function}) for the P-function allow us to overcome this problem and to derive versions of the Ehrenfest theorem in terms of Wick and Anti-Wick symbols and star products, namely:
\begin{equation*}
    \begin{aligned}
        \frac{d}{dt}\langle\hat A\rangle 
        & = -\frac{i}{\hbar}\langle\{\{A_{(a)W}\,,\,H_{(a)W}\}\}_{*_{(a)W}}\rangle\,.\label{ehrenfest_wick}
    \end{aligned}
\end{equation*} \\ \par
 
Consider the expectation value of an observable $\hat A$ written as 
via the Q-function,
\begin{equation*}
    \begin{aligned}
        \langle \hat A\rangle = \int~d^2 \alpha~A_{aW}(\alpha,\alpha^*)\cdot Q(\alpha,\alpha^*,t)\,.
    \end{aligned}
\end{equation*}
Taking the time derivative yields
\begin{equation*}
    \begin{aligned}
        \frac{d}{dt} \langle \hat A\rangle = \int~d^2\alpha~A_{aW}\cdot \frac{\partial}{\partial t}Q(t)\,.
    \end{aligned}
\end{equation*}
 
Using Eq.\ (\ref{dynamical eqn for Q-function}) we get
\begin{equation*}\resizebox{\columnwidth}{!}{%
    $\displaystyle
    \begin{aligned}
        \frac{d}{dt} \langle A\rangle = &\int~d^2\alpha~A_{aW}\\
        &\Bigl\{-\frac{i}{\hbar}\sum_{n=1}^{\infty}
    \frac{1}{n!}
    \;\left[\partial_{\alpha^*}^{n}\!\Bigl(
      \partial_{\alpha}^{n}\,H_{aW}\;\cdot Q
    \Bigr)-\partial_{\alpha}^{n}\!\Bigl(
      \partial_{\alpha^*}^{n}\,H_{aW}\;\cdot Q
    \Bigr)\right]\Bigr\} \,,\\
    = & -\frac{i}{\hbar} \sum_{n=1}^\infty\frac{1}{n!} \int ~d^2\alpha~A_{aW}\\
    &~~~~~~~~~~~~~~~~\left[\partial_{\alpha^*}^{n}\!\Bigl(
      \partial_{\alpha}^{n}\,H_{aW}\;\cdot Q
    \Bigr)-\partial_{\alpha}^{n}\!\Bigl(
      \partial_{\alpha^*}^{n}\,H_{aW}\;\cdot Q
    \Bigr)\right]\,.
    \end{aligned}$}
\end{equation*}
Next, using Eq.\ (\ref{Shifting the derivatives to surface derivatives}) (with $f=A_{aW}$, $g=\partial_\alpha^nH_{aW}\,\cdot Q$) we obtain
\begin{equation*}
\begin{aligned}A_{aW}&\cdot\partial_{\alpha^*}^n(\partial_\alpha^nH_{aW}\,\cdot Q)\\
    =&\sum_{k=0}^n\binom{n}{k}\left(-1\right)^{\textcolor{black}{n-k}}\\
    &~~\partial_{\alpha^*}^k\left(\partial_{\alpha^*}^{n-k}A_{aW}\, \cdot (\partial_\alpha^nH_{aW}\,\cdot Q)\right)\,.
\end{aligned}
\end{equation*}
This gives
\begin{equation*}
    \begin{aligned}
        \int &~d^2\alpha~A_{aW}\cdot\partial_{\alpha^*}^{n}\!\Bigl(
      \partial_{\alpha}^{n}\,H_{aW}\;\cdot Q
    \Bigr) \\
    & = \int~d^2\alpha~\sum_{k=0}^n \binom{n}{k}\left(-1\right)^{\textcolor{black}{n-k}}\,\\
    &~~~~~~~~~~~~~~~~~~~~~~~~\partial_{\alpha^*}^k\left(\partial_{\alpha^*}^{n-k}A_{aW}\, \cdot (\partial_\alpha^nH_{aW}\,\cdot Q)\right)\,,\\
    & = \int~d^2\alpha~ (-1)^n(\partial_{\alpha^*}A_{aW}\cdot \partial_\alpha^nH_{aW})\,\cdot Q \\&~~~~+ \int~d^2\alpha~\sum_{k=1}^n \binom{n}{k}\left(-1\right)^{\textcolor{black}{n-k}}\,\\
    &~~~~~~~~~~~~~~~~~~~~~~\partial_{\alpha^*}^k\left(\partial_{\alpha^*}^{n-k}A_{aW}\, \cdot (\partial_\alpha^nH_{aW}\,\cdot Q)\right).
    \end{aligned}
\end{equation*}
The second summation term above in the last line is a collection of surface terms, all of which vanish at the phase space boundary since the Q-function vanishes at the phase space boundary. Accordingly,
\begin{equation*}
    \begin{aligned}
    \int &~d^2\alpha~A_{aW}\cdot\partial_{\alpha^*}^{n}\!\Bigl(
      \partial_{\alpha}^{n}\,H_{aW}\;\cdot Q
    \Bigr)  \\
    & = \int~d^2\alpha~ (-1)^n(\partial_{\alpha^*}A_{aW}\cdot \partial_\alpha^nH_{aW})\,\cdot Q
    \end{aligned}
\end{equation*}
and, analogously,
\begin{equation*}
    \begin{aligned}
     \int &~d^2\alpha~A_{aW}\cdot\partial_{\alpha}^{n}\!\Bigl(
      \partial_{\alpha^*}^{n}\,H_{aW}\;\cdot Q
    \Bigr)  \\
    & = \int~d^2\alpha~ (-1)^n(\partial_{\alpha}A_{aW}\cdot \partial_{\alpha^*}^n H_{aW})\,\cdot Q
    \end{aligned}
\end{equation*}
Using these results, we write down the time derivative of the expectation value of the observable in the following way,
\begin{equation*}\resizebox{\columnwidth}{!}{%
    $\displaystyle
    \begin{aligned}
        \frac{d}{dt}\langle \hat A \rangle  =& -\frac{i}{\hbar} \sum_{n=1}^\infty\frac{(-1)^n}{n!} \int ~d^2\alpha~\\
        &~~\Bigl[\partial_{\alpha^*}A_{aW}\cdot \partial_\alpha^nH_{aW}-\partial_{\alpha}A_{aW}\cdot \partial_{\alpha^*}^n H_{aW}\Bigr]\,\cdot Q \,,\\
        =& -\frac{i}{\hbar} \int~d^2\alpha~\left\{\sum_{n=1}^\infty\frac{(-1)^n}{n!}\right.\\
        &~~~~\left.\Bigl[\partial_{\alpha^*}A_{aW}\cdot \partial_\alpha^nH_{aW}-\partial_{\alpha}A_{aW}\cdot \partial_{\alpha^*}^n H_{aW}\Bigr]\right\}\, \cdot Q\,,\\
        =&  -\frac{i}{\hbar} \int~d^2\alpha~\\
        &~~~~~~\Bigl\{A_{aW} e^{-\overset{\xleftarrow{}}{\partial_{\alpha^*}}\overset{\xrightarrow{}}{\partial_{\alpha}}} H_{aW}- H_{aW}e^{-\overset{\xleftarrow{}}{\partial_{\alpha^*}}\overset{\xrightarrow{}}{\partial_{\alpha}}}A_{aW}\Bigl\}\, \cdot Q\,,\\
        =& -\frac{i}{\hbar} \int~d^2\alpha~\\
        &~~~~~~\Bigl( A_{aW}*_{aW}H_{aW} - H_{aW}*_{aW}A_{aW}\Bigr)\,\cdot Q\,,\\
        =&-\frac{i}{\hbar}\int~d^2\alpha~\Bigl(\{\{A_{aW}, H_{aW}\}\}_{*_{aW}}\Bigr)\,\cdot Q\,,\\
        =& -\frac{i}{\hbar} \langle \{\{A_{aW}, H_{aW}\}\}_{*_{aW}} \rangle\,,
    \end{aligned}$}
\end{equation*}
which is exactly the announced result.
\\ \par

The calculation of the Ehrenfest theorem for the Wick symbols can be done analogously. The only difference is that for the case of Wick symbols the dynamical evolution of the P-function contains a sum over $\frac{(-1)^n}{n!}$ while the dynamical equation for the Q-function contains a sum over $\frac{1}{n!}$. This difference is relevant for introducing the corresponding star products when deriving the respective Ehrenfest theorem.

We conclude that the Ehrenfest theorem can be written in its fully general form as
\begin{equation*}
    \frac{d}{dt}\langle A \rangle = -\frac{i}{\hbar} \, \langle \{\{A_o,H_o\}\}_{*_o},
\end{equation*}
where $o \in \{W,aW, Weyl\}$ is the shorthand for the different symbols (Weyl, Wick and Anti-Wick) of the observable operators and the corresponding deformed star algebra associated with the respective dequantization type. 


\section{Anharmonic Oscillator}\label{Anharmonic oscillator}

Our results for the time evolution of the Husimi $Q-$function are in apparent conflict with those obtained in Milburn's study of the quantum anharmonic oscillator \cite{Milburn_article}. Notably, according to the results derived here (see Eqs.\ (\ref{drift-diffusion}) and (\ref{Fokker-Planck})), for the Hamiltonians of the harmonic and anharmonic oscillator, the time evolution of the Husimi function follows a Fokker-Planck equation with a drift-term $(A_q,A_p)=(\partial_pH_{aW},-\partial_qH_{aW})$ which, if Anti-Wick quantization is used, corresponds exactly to the classical result, namely, the Liouville term derived from Hamilton's equations. According to Milburn, in contrast, the drift term in the Fokker-Planck-like equation for the Q-function has a non-classical correction term (Eq.\ (30) of \cite{Milburn_article}). Here we show that there is no conflict and that Milburn's ``quantum correction'' to the drift term is an artifact of the fact that Milburn uses a quantum Hamiltonian that is not generated from the classical Hamiltonian via Anti-Wick quantization and, hence, not a natural fit with using the Husimi function as the phase space distribution function. 

The classical Hamiltonian for the anharmonic oscillator is
\begin{equation}
\begin{aligned}
    H&=\hbar \omega_0 \left(|\alpha|^2+\mu|\alpha|^4\right)\\
    &=H_0+\frac{\mu}{\hbar\omega_0}H_0^2\,, \label{anharmonic}
\end{aligned}
\end{equation}
where $\mu$ is a positive anharmonicity parameter. If we treat this classical Hamiltonian as an Anti-Wick symbol, as is natural when using the Husimi Q-function as our phase space distribution function, then, using the $n=1$-term in Eq.\ (\ref{dynamical eqn for Q-function}), the drift term of the flow of the Husimi function is given by
\begin{equation}
\begin{aligned}
        A_{\alpha^*}&=\partial_\alpha H_{aW}=\hbar\omega_0\left(\alpha^*+2\mu\alpha^*|\alpha|^2\right)\,,\\
        A_{\alpha}&=-\partial_{\alpha^*}H_{aW}=-\hbar\omega_0\left(\alpha+2\mu\alpha|\alpha|^2\right)\,.
\end{aligned}
\end{equation}
This coincides exactly with the classical result given by Hamilton's equations.

Milburn, in contrast, identifies a discrepancy between the classical result and the quantum result, in apparent conflict with our finding. His calculations compare the time evolution of the classical phase space probability density in a rotating frame, where the solution to the corresponding harmonic oscillator is constant, to the time evolution of the Husimi function in ther interaction picture. In the rotating frame, the time evolution of the classical phase space probability density $Q_{class}(t)$ is given by
\begin{equation}\label{Milburn classical probability dynamical eqn}%
    \begin{aligned}
        \frac{\partial Q_{class}}{\partial \tau}
        =&2i\mu\left[\partial_\alpha(\alpha |\alpha|^2 Q_{class}) - \partial_{\alpha^*}(\alpha^*|\alpha|^2Q_{class})\right]\,,
    \end{aligned}
\end{equation}
where $\tau$ is the dimensionless time $\tau=\omega_0t$ (see Eq.\ (14) of \cite{Milburn_article}). \\\par 

For the corresponding quantum calculation, Milburn uses the quantum Hamiltonian operator
\begin{equation}\label{Milburn_Hamiltonian}
\begin{aligned}
    \hat H_{\rm Milburn}&=\hat H_0+\frac{\mu}{\hbar\omega_0}\hat H_0^2\\
    &=\hbar \omega_0[(1+\mu)\hat a^\dagger \hat a+\mu(\hat a^\dagger \hat a)^2]+const.\,.\end{aligned}
\end{equation}
To derive the time evolution of the Q-function, Milburn uses the interaction picture with the interaction Hamiltonian $\hat H_I=\hbar \omega_o\mu (\hat a^\dagger \hat a)^2$. Using this interaction Hamiltonian, we re-derive the discrepancy between the quantum drift and the classical Liouville term as follows. Since the interaction Hamiltonian is a product of Wick quantized operators, i.e., $(\hat a^\dagger \hat a)^2$, using the Wick symbol algebra we have $(\hat a^\dagger \hat a)^2=\mathcal{Q}^W(\alpha^*\alpha *_W \alpha^*\alpha)=\mathcal{Q}^W(\alpha^{*2}\alpha^2+\alpha^*\alpha)$. So the Wick symbol of the interaction Hamiltonian is $H_{I,W}=\hbar \omega_o\mu(|\alpha|^4+|\alpha|^2)$. The $\hbar \omega_0\mu|\alpha|^4$ term corresponds to the classical effective Hamiltonian in the rotated frame. The anti-Wick symbol of the Hamiltonian can be obtained from its Wick symbol by $H_{I,aW}=e^{-\partial_\alpha\partial_{\alpha^*}}H_{I,W}=\hbar \omega_o\mu(\alpha^{*2}\alpha^2-3\alpha^*\alpha+1)$. Using this anti-Wick symbol in Eq. (\ref{dynamical eqn for Q-function}), we obtain:
\begin{equation}
    \begin{aligned}
        \partial_\tau Q  =& \frac{i}{\hbar}(\hbar \mu)\left[\partial_\alpha((2\alpha |\alpha|^2-3\alpha)Q) \right.\\
        &\left. - \partial_{\alpha^*}((2\alpha^*|\alpha|^2-3\alpha^*)Q)\right.\\
        &\left. +\frac{1}{2}\left\{\partial_\alpha^2(2\alpha^2Q)-\partial_{\alpha^*}^2(2\alpha^{*2}Q)\right\}\right]\\
        =&i\mu \alpha(1+2|\alpha|^2)\partial_\alpha Q - i\mu \alpha^*(1+2|\alpha|^2)\partial_{\alpha^*} Q \,\\
    &+i\mu\alpha^2 \partial_\alpha^2Q -i\mu\alpha^{*2} \partial_{\alpha^*}^2Q\,.
    \end{aligned}
\end{equation}
The result in the last two lines is exactly the one obtained by Milburn obtained as Eq. (30) in \cite{Milburn_article}.
The most obvious difference between this result and the classical rotating frame result Eq.\ (\ref{Milburn classical probability dynamical eqn}) are the diffusion terms in the last line of this expression.
However, there is also a discrepancy between the quantum and classical drift terms, which stems from the terms $i\mu\alpha$ and $i\mu\alpha^*$ in the second-to-last line, which cannot be found in Eq.\ (\ref{Milburn classical probability dynamical eqn})

However, this discrepancy is an artifact of the specific quantum Hamiltonian Eq.\ (\ref{Milburn_Hamiltonian}) used by Milburn. That Hamiltonian is not obtained from the classical Hamiltonian Eq.\ (\ref{anharmonic}) by Anti-Wick quantization and, hence, arguably not the suitable one to use when computing the time evolution of the Husimi function specifically. Quantizing the classical Hamiltonian Eq.\ (\ref{anharmonic}) via Anti-Wick quantization, i.e. using anti-normal ordering of creation and annihilation operators, yields the quantum Hamiltonian 
\begin{equation*}
\begin{aligned}
    \hat H
    &=\hbar \omega_0[\hat a a^\dagger+\mu(\hat a \hat a\hat a^\dagger\hat a^\dagger)]\,\\
    &=\hbar \omega_0[(1+3\mu)\hat a^\dagger \hat a+\mu(\hat a^\dagger \hat a)^2]+const.\label{AW_Hamiltonian}
\end{aligned}
\end{equation*}
Trivially, due to how this Hamiltonian was constructed, its Anti-Wick symbol is $H$ from Eq.\ (\ref{anharmonic}). When we plug this $H$ into the time evolution equation for the Husimi function derived earlier, Eq. (\ref{dynamical eqn for Q-function}), the first order term matches the classical result Eq. (\ref{Milburn classical probability dynamical eqn}) exactly:
\begin{equation*}\resizebox{\columnwidth}{!}{
   $\displaystyle
\begin{aligned}
    \partial_\tau Q&=-i\mu\left[\partial_{\alpha^*}^{}\Bigl(
      \partial_{\alpha}^{}|\alpha|^4\cdot Q
    \Bigr)-\partial_{\alpha}^{}\Bigl(
       \partial_{\alpha^*}^{}|\alpha|^4\cdot Q
    \Bigr)\right]\\
    &~~~~~~+ \frac{i\mu}{2}\left[\partial_{\alpha}^2(\partial_{\alpha^*}^2|\alpha|^4\cdot Q)+\partial_{{\alpha}^{*}}^2(\partial_\alpha^2|\alpha|^4\cdot Q)\right]  \\    
    &=-2i\mu\left[\partial_{\alpha^*}^{}\Bigl(
      \alpha^*|\alpha|^2\cdot Q
    \Bigr)-\partial_{\alpha}^{}\Bigl(
       \alpha|\alpha|^2\cdot Q
    \Bigr)\right] \\
    &~~~~~~+i\mu \left[\partial_{\alpha}^2(\alpha^2\cdot Q)+\partial_{\alpha^*}^2(\alpha^{*2}\cdot Q)\right]\,.
\end{aligned}$}
\end{equation*}
There is thus no quantum contribution to the drift beyond the classical Liouville term if the appropriate -- Anti-Wick -- quantization scheme is used when comparing the evolution of the Husimi function to its classical analogue. The solution to the above equation for the initial state as a coherent state, i.e., $Q(\alpha,\alpha^*,0)=\exp(-|\alpha-\alpha_0|^2)$ is 
\begin{equation}
    Q(\alpha, \alpha^*, \tau)=\exp\left(-|\alpha|^2-|\alpha_0|^2\right)|S(\alpha,\alpha^*, \tau)|^2\,,
\end{equation}
where 
\begin{equation}
    S(\alpha, \alpha^*, \tau)=\sum_{n=0}^\infty\frac{(\alpha_0^*\alpha)^n}{n!}\exp\left(i\mu\tau[n(n-1)]\right)\,.
\end{equation}
In the limit $\hbar \to 0$, the above expression reduces to the corresponding classical expression, obtained by solving equation (\ref{Milburn classical probability dynamical eqn}). 

\section{Conclusion}\label{conclusion}
In this paper we have derived time evolution equations for the Husimi Q- and Glauber-Sudarshan P-functions in terms of complementary Hamiltonian symbols, i.e. Anti-Wick Hamiltonian symbol for the Husimi function and Wick Hamiltonian symbol for the Glauber-Sudarshan function. This has enabled us to derive a unified formulation of Ehrenfest's theorem that applies to Weyl, Wick, and Anti-Wick symbols alike. While these Ehrenfest-type results do not qualitatively go beyond standard correspondence results for the quantum-classical relation, they demonstrate that the complementary-symbol formalism provides a coherent framework across Wick, Anti-Wick, and Weyl representations. This adds to the internal consistency of phase-space approaches and may be useful as a starting point for further developments. Finally, we have demonstrated that the appearance of a quantum contribution to the drift, over and above the Liouvillian term, in the time evolution of the Husimi function of the anharmonic oscillator obtained by Milburn in \cite{Milburn_article}, together with the dynamical consequences of that term, is an artifact of using a quantum Hamiltonian not obtained by Anti-Wick quantization. This observation underscores the importance of using a complementary quantization scheme when studying the time evolution of phase space distribution functions, notably when comparing with the classical case.

\section*{Acknowledgments}
We would like to thank Gerard Milburn for helpful feedback. We used OpenAI's GPT 5 for copyediting.

This research was funded by the Netherlands Organization for Scientific Research (NWO), project VI.Vidi.211.088.

\appendix
\section{}\label{Proof of identity used to shift derivatives}
In this section we prove the following result,
\begin{equation}\label{*}
\bigl(\partial_\alpha^k f\bigr)\partial_{\alpha^*}^n g
=\sum_{j=0}^n \binom{n}{j}\left(-1\right)^{n-j} 
\partial_{\alpha^*}^j\!\left(\partial_{\alpha^*}^{\,n-j}\bigl(\partial_\alpha^k f\bigr) g\right)\,, 
\end{equation}
by induction. \\
\par 
\begin{proof}

\emph{Base case.} For \(n=0\) we get 
\[
\bigl(\partial_\alpha^k f\bigr) g 
= \binom{0}{0}(-1)^0 \partial_{\alpha^*}^0\!\left(\partial_{\alpha^*}^0 \bigl(\partial_\alpha^k f\bigr) g\right)
= \bigl(\partial_\alpha^k f\bigr) g,
\]
which is true.\\

\emph{Inductive step.} Assume equation (\ref{*}) holds for some \(n\). We prove that it also holds for \(n+1\).
Using the Leibniz rule, we write
\[
\bigl(\partial_\alpha^k f\bigr)\partial_{\alpha^*}^{\,n+1} g
= \partial_{\alpha^*}\!\Bigl((\partial_\alpha^k f)\,\partial_{\alpha^*}^n g\Bigr) 
- \bigl(\partial_{\alpha^*}\partial_\alpha^k f\bigr)\,\partial_{\alpha^*}^n g.
\]
Apply equation (\ref{*}) to both \(\partial_\alpha^k f\) and to \(\partial_{\alpha^*}\partial_\alpha^k f\) (the latter is allowed since these derivatives have the required regularity):
\[\resizebox{\columnwidth}{!}{
    $\displaystyle
\begin{aligned}
\partial_{\alpha^*}\!\Bigl((\partial_\alpha^k f)\,\partial_{\alpha^*}^n g\Bigr)
&= \sum_{j=0}^{n}\binom{n}{j}(-1)^{\,n-j}\,
\partial_{\alpha^*}^{\,j+1}\!\Bigl(\partial_{\alpha^*}^{\,n-j}(\partial_\alpha^k f)\,g\Bigr),\\[2mm]
\bigl(\partial_{\alpha^*}\partial_\alpha^k f\bigr)\,\partial_{\alpha^*}^n g
&= \sum_{j=0}^{n}\binom{n}{j}(-1)^{\,n-j}\,
\partial_{\alpha^*}^{\,j}\!\Bigl(\partial_{\alpha^*}^{\,n-j+1}(\partial_\alpha^k f)\,g\Bigr).
\end{aligned}$}
\]
Subtracting the two sums gives
\[\resizebox{\columnwidth}{!}{
    $\displaystyle
\begin{aligned}
(\partial_\alpha^k f)\,\partial_{\alpha^*}^{\,n+1} g
= \sum_{j=0}^{n}&\binom{n}{j}(-1)^{\,n-j}\\
&\Bigl[
\partial_{\alpha^*}^{\,j+1}\!\Bigl(\partial_{\alpha^*}^{\,n-j}(\partial_\alpha^k f)\,g\Bigr)
-\partial_{\alpha^*}^{\,j}\!\Bigl(\partial_{\alpha^*}^{\,n-j+1}(\partial_\alpha^k f)\,g\Bigr)\Bigr].
\end{aligned}$}
\]
Reindex the first sum by \(m=j+1\) (so \(m=1,\dots,n+1\)) while leaving the second sum as is (\(m=0,\dots,n\)),

\[
\begin{aligned}
S_1&=\sum_{m=1}^{n+1}\binom{n}{m-1}(-1)^{\,n+1-m}\,
\partial_{\alpha^*}^{\,m}\!\Bigl(\partial_{\alpha^*}^{\,n+1-m}(\partial_\alpha^k f)\,g\Bigr)\,,\\
S_2&=\sum_{m=0}^{n}\binom{n}{m}(-1)^{\,n-m}\,
\partial_{\alpha^*}^{\,m}\!\Bigl(\partial_{\alpha^*}^{\,n+1-m}(\partial_\alpha^k f)\,g\Bigr).
\end{aligned}
\]
Hence \((\partial_\alpha^k f)\partial_{\alpha^*}^{\,n+1}g=S_1-S_2\). For each fixed \(m\) the contribution to
\(S_1-S_2\) is: for \(m=0\) only \(S_2\) contributes with coefficient 
\[
 -\binom{n}{0}(-1)^{n}=(-1)^{n+1}\binom{n}{0}\,,
\]
for \(m=n+1\) only \(S_1\) contributes with coefficient
\[
 \binom{n}{n}(-1)^{0}=\binom{n}{n}\,,
\]
and for \(1\le m\le n\) the coefficient is
\begin{equation*}
\begin{aligned}
\binom{n}{m-1}(-1)^{\,n+1-m}&-\binom{n}{m}(-1)^{\,n-m}\\
&= (-1)^{\,n+1-m}\Bigl(\binom{n}{m-1}+\binom{n}{m}\Bigr).
\end{aligned}
\end{equation*}
Combining termwise and using the binomial identity
\[
\binom{n}{m-1}+\binom{n}{m}=\binom{n+1}{m},
\]
yields
\[\begin{aligned}
(\partial_\alpha^k f)\,\partial_{\alpha^*}^{\,n+1} g
= \sum_{m=0}^{n+1}\binom{n+1}{m}(-1)^{\,n+1-m}\,\\
\partial_{\alpha^*}^{\,m}\!\Bigl(\partial_{\alpha^*}^{\,n+1-m}(\partial_\alpha^k f)\,g\Bigr),
\end{aligned}\]
which is precisely equation (\ref{*}) for \(n+1\). This completes the induction.
\begin{remark}
The argument requires only that $\partial_{\alpha^*}$ is a derivation (Leibniz rule) and that $\partial_\alpha^k f$ along with the derivatives $\partial_{\alpha^*}(\partial_\alpha^k f), \partial_{\alpha^*}^2(\partial_\alpha^k f),\dots$ are regular enough.
\end{remark} 
\end{proof}

\section{}\label{Appendix B}
In this section we show the derivation for Eq. (\ref{Wick star bracket in terms of anti-Wick symbols}). We start by considering the expression,
\begin{equation*}
\resizebox{\columnwidth}{!}{%
    $\displaystyle
\begin{aligned}
    \{\{f,g\}\}_{*_{W}} & = f*_{W}g-g*_{W}f\,,\\
    & = \sum_{k=1}^\infty \frac{1}{k!} \left[\sum_{m=1}^k (-1)^{k-m}\,\binom{k}{m}\right. \,\\
    & \,\left. \left\{\partial_{\alpha^*}^m\left(\partial_{\alpha^*}^{k-m} \partial_{\alpha}^k f\, g\right)-\partial_{\alpha}^m\left( \partial_{\alpha}^{k-m} \partial_{\alpha^*}^kf\, g\right)\right\}\right]\,.
\end{aligned}
$}
\end{equation*}

Take,
\begin{equation*}
    \sum_{k=1}^\infty \frac{1}{k!} \left[\sum_{m=1}^k (-1)^{k-m}\binom{k}{m} \,  \partial_{\alpha^*}^m\left( \partial_{\alpha^*}^{k-m} \partial_{\alpha}^kf\, g\right)\right]\,, 
\end{equation*}
and perform the linear change of summation variables,
\begin{equation*}
\begin{aligned}
   n &= m,\quad\\
   t &= k - m,\quad\\
   k &= n + t,
\end{aligned}\end{equation*}
so that \(n\ge1\) and \(t\ge0\).  Under this re-indexing, the combinatorial factors become,
\begin{equation*}
\begin{aligned}
   \binom{k}{m} &= \binom{n+t}{n},\quad
   (-1)^{k-m} = (-1)^t\,,\\ 
   \frac1{k!}\Bigl(\tfrac\hbar2\Bigr)^k
   &= \frac1{(n+t)!}\Bigl(\tfrac\hbar2\Bigr)^{n+t}.
\end{aligned}
\end{equation*}
The summation can now be rewritten as,
\begin{equation*}
\begin{aligned}
  \sum_{k=1}^\infty & \frac{1}{k!} \left[\sum_{m=1}^k (-1)^{k-m}\binom{k}{m} \, \partial_{\alpha^*}^m\left( \partial_{\alpha^*}^{k-m} \partial_{\alpha}^kf\, g\right)\right]\,\\
   = & \sum_{n=1}^{\infty}\sum_{t=0}^{\infty}
    \frac{1}{(n+t)!}
    \binom{n+t}{n}
    (-1)^t
    \;\\
    &~~~~~~~~~~~~~~~~~~~~~~\partial_{\alpha^*}^{n}\!\Bigl((\partial_{\alpha^*}^{t}\,\partial_{\alpha}^{\,n+t}f)\,g\Bigr).
\end{aligned}
\end{equation*}

The factorial/binomial combination simplifies to
\begin{equation*}
  \frac1{(n+t)!}\binom{n+t}{n}
  = \frac1{n!\,t!},
\end{equation*}
so that the right hand side becomes,
\begin{equation*}
  \sum_{n=1}^{\infty}\sum_{t=0}^{\infty}
    \frac{1}{n!}
    \;\frac{(-1)^t}{t!}
    \;\partial_{\alpha^*}^{n}\!\Bigl((\partial_{\alpha^*}^{t}\,\partial_{\alpha}^{\,n+t}f)\,g\Bigr).
\end{equation*}

Since \(\partial_{\alpha}\) and \(\partial_{\alpha^*}\) commute, we may group the t-derivatives of \(\partial_{\alpha}\) onto \(f\) first, then apply the remaining n-derivatives:
\begin{equation*}
    \begin{aligned}
  (\partial_{\alpha}^{\,n+t}f)\,g
  &= \partial_{\alpha}^n\!\bigl(\partial_{\alpha}^t f\bigr)\;\cdot g\,,
  \quad\\
  \partial_{\alpha^*}^{n}\!\bigl((\partial_{\alpha^*}^t\,\partial_{\alpha}^{\,n+t}f)\,g\bigr)
  &= \partial_{\alpha^*}^n\!\Bigl(\partial_{\alpha}^n\!\bigl(\partial_{\alpha^*}^t\partial_{\alpha}^t f\bigr)\;\cdot g\Bigr).
  \end{aligned}
\end{equation*}
Substituting back gives
\begin{equation*}
  \sum_{n=1}^{\infty}\sum_{t=0}^{\infty}
    \frac{1}{n!}
    \;\partial_{\alpha^*}^{n}\!\Bigl[
      \partial_{\alpha}^{n}\!\bigl(\tfrac{(-1)^t}{t!}(\partial_{\alpha^*}\partial_{\alpha}^t f)\bigr)\;\cdot g
    \Bigr].
\end{equation*}
Finally the summation can be written as,  
\begin{equation*}
\begin{aligned}
    \sum_{k=1}^\infty & \frac{1}{k!} \left[\sum_{m=1}^k \binom{k}{m}(-1)^{k-m} \,  \partial_{\alpha^*}^m\left( \partial_{\alpha^*}^{k-m} \partial_{\alpha}^kf\, g\right)\right]\\
     =& \sum_{n=1}^{\infty}
    \frac{1}{n!}
    \;\partial_{\alpha^*}^{n}\!\Bigl[
      \partial_{\alpha}^{n}\!\bigl(\sum_{t=0}^{\infty}\tfrac{(-1)^t}{t!}(\partial_{\alpha^*}\partial_{\alpha})^t f)\bigr)\;\cdot g
    \Bigr]\,,\\
    =&\sum_{n=1}^{\infty}
    \frac{1}{n!}\Bigl(\tfrac\hbar2\Bigr)^{n}
    \;\partial_{\alpha^*}^{n}\!\Bigl[
      \partial_{\alpha}^{n}\!\bigl(e^{\frac{-\hbar}{2}\partial_{\alpha^*}\partial_{\alpha}}f\bigr)\;\cdot g
    \Bigr]\,.
\end{aligned}
\end{equation*}

We note that for Wick symbols $f$, the Anti-Wick symbol is given as $\bar f=e^{-\partial_{\alpha^*}\partial_{\alpha}}f$ (referred to as the inverse Berezin transform, since formally the Berezin transform of any function $f$ is given as $e^{\partial_{\alpha^*} \partial_{\alpha}}f$ \cite{Ali:2004ft}). Hence we can write,
\begin{equation*}
\begin{aligned}
    \sum_{k=1}^\infty & \frac{1}{k!} \left[\sum_{m=1}^k \binom{k}{m}(-1)^{k-m} \, \partial_{\alpha^*}^m\left( \partial_{\alpha^*}^{k-m} \partial_{\alpha}^kf\, g\right)\right] \\
    =& \sum_{n=1}^{\infty}
    \frac{1}{n!}
    \;\partial_{\alpha^*}^{n}\!\Bigl[
      \partial_{\alpha}^{n}\,\bar f\;\cdot g
    \Bigr]\,.
\end{aligned}
\end{equation*}
Following this we get,
\begin{equation*}
\resizebox{\columnwidth}{!}{%
    $\displaystyle
\begin{aligned}
    \{\{f,g\}\}_{*_{W}}&=\sum_{n=1}^{\infty}
    \frac{1}{n!}
    \;\left[\partial_{\alpha^*}^{n}\!\Bigl(
      \partial_{\alpha}^{n}\,\bar f\;\cdot g
    \Bigr)-\partial_{\alpha}^{n}\!\Bigl(
      \bar \partial_{\alpha^*}^{n}\,\bar f\;\cdot g
    \Bigr)\right]\,.
\end{aligned}
$}
\end{equation*}
\\ \par

\vspace{2cm}

\bibliographystyle{apsrev4-2}
\bibliography{bibliography}

\end{document}